\documentclass[prb,a4paper,twocolumn,showpacs,superscriptaddress,floatfix]{revtex4}
\usepackage{times}

\usepackage{graphicx}
\usepackage{amsmath}
\usepackage{amssymb}
\usepackage{bm}
\usepackage{color}
\usepackage[normalem]{ulem}
\newcommand{\be}{\begin{equation}}
\newcommand{\ee}{\end{equation}}

\begin{document}

\title{Generalized correlation functions for conductance fluctuations and the mesoscopic spin Hall effect}

\author{J. G. G. S. Ramos}
\affiliation{Instituto de Estudos Avan\c cados, Universidade de S\~{a}o Paulo, C.P.\ 72012, 05508-970 S\~{a}o Paulo - SP, Brazil}
\affiliation{Instituto de F\'{\i}sica, Universidade de S\~{a}o Paulo, C.P.\ 66318, 05314-970 S\~{a}o Paulo - SP, Brazil}
\affiliation{Departamento de Ci\^encias Exatas, Universidade Federal da Para\'{\i}ba, 58297-000, Rio Tinto - PB, Brazil}

\author{A. L. R. Barbosa}
\affiliation{UAEADTec and P\'os-Gradua\c c\~ao em F\'isica Aplicada, Universidade Federal Rural de Pernambuco, 52171-900 Recife - PE, Brazil}

\author{D. Bazeia}
\affiliation{Departamento de F\'{\i}sica, Universidade Federal da Para\'{\i}ba, 58051-970 Jo\~{a}o Pessoa - PB, Brazil}
\affiliation{Instituto de F\'{\i}sica, Universidade de S\~{a}o Paulo, C.P.\ 66318, 05314-970 S\~{a}o Paulo - SP, Brazil}

\author{M. S. Hussein}
\affiliation{Instituto de Estudos Avan\c cados, Universidade de S\~{a}o Paulo, C.P.\ 72012, 05508-970 S\~{a}o Paulo - SP, Brazil}
\affiliation{Instituto de F\'{\i}sica, Universidade de S\~{a}o Paulo, C.P.\ 66318, 05314-970 S\~{a}o Paulo - SP, Brazil}

\author{C. H. Lewenkopf}
\affiliation{Instituto de F\'{\i}sica, Universidade Federal Fluminense, 24210-346 Niter\'{o}i - RJ, Brazil}

\date{\today}

\begin{abstract}
We study the spin-Hall conductance fluctuations in ballistic mesoscopic systems. We obtain universal expressions for the spin and charge current fluctuations, cast in terms of current-current autocorrelation functions. We show that the latter  are conveniently parametrized as deformed Lorentzian shape lines, functions of an external applied magnetic field and the Fermi energy.  We find that the charge current fluctuations show quite unique statistical features at the symplectic-unitary crossover regime. Our findings are based on an evaluation of the generalized transmission coefficients correlation functions within the stub model and are amenable to experimental test.
\end{abstract}

\pacs{05.45.Yv, 03.75.Lm, 42.65.Tg}

\maketitle

\section{Introduction}
\label{sec:introduction}

The discovery of the spin Hall effect (SHE) \cite{mote1,mote2,mote3,mote4} in both metal and semiconductor structures has opened an important new possibility to control the effects of non-equilibrium spin accumulation. \cite{jungwith12} The basic idea underlying the SHE is to generate spin currents transverse to the longitudinal transport of charge by creating an imbalance between the spin up and spin down states. \cite{mote5,mote6}

The detection of spin-Hall conductance fluctuations is a major goal of semiconductor spintronics. \cite{revisao1} It is, however, a hard endeavor. The main reason is the difficulty to efficiently connect ferromagnets leads to two-dimensional semiconductor structures. \cite{revisao2} For ballistic systems coupled metallic leads, it is in principle possible to detect the signal when scattering by impurities induce a separation of the spin states.

Through this mechanism, universal spin-Hall conductance fluctuations (USCF) can lead to accumulation of spin at the electron reservoirs. The USCF appear in the transverse current measured in multi-terminal devices in the presence of a sufficiently large magnetic field. \cite{mote5}
Signals of the spin accumulation can be inferred, for instance, from time-dependent fluctuations of the spectral currents (noise power) \cite{nos1} or from the analysis of universal conductance fluctuations. \cite{simulacao,jacquod1,Nazarov2007,Krich2008} Spin-Hall conductance fluctuations have been theoretically studied for mesoscopic systems in the diffusive \cite{simulacao} as well as in the ballistic regime. \cite{jacquod1} In the absence of both spin rotation symmetry and magnetic field, these studies predict  universal spin-Hall conductance fluctuations with a root mean square amplitude of about  $0.18 (e/ 4 \pi)$.

So far, a direct detection of spin-Hall currents by analyzing transverse current fluctuations has not been made. In this paper, we propose an alternative way to infer  spin-Hall conductance fluctuations, based on the universal relation between spin and charge current fluctuations in chaotic quantum dots. We find that the change and spin current-current correlation functions show a quite unique dependence on the ratio of open modes between transversal and longitudinal terminals. This dependence allows one to infer the magnitude of the spin current. From the technical point of view, we adapt the diagrammatic technique developed to describe the electronic transport in two-terminal chaotic quantum dots in the presence of a spin-orbit interaction, at the symplectic-unitary corssover, \cite{brouwer2,Cremers03} to the case of multi-terminal spin resolved currents.

The paper is organized as follows. In Sec.\ \ref{sec:theory} we review the Landauer-B\"uttiker approach used to calculate the multi-terminal charge and spin currents. Next, in Sec.~\ref{sec:disorder}, we present the diagrammatic theory we employ to calculate the universal spin Hall current fluctuations. The phenomenological implications of our findings are discussed in Sec.~\ref{sec:results}. Finally, in Sec.~\ref{sec:conclusions}, we present our conclusions.

\section{Theoretical framework}
\label{sec:theory}

In this section we review the scattering matrix formalism that describes the spin Hall effect in ballistic conductors. We follow the approach put forward in Refs.~\onlinecite{jacquod1,Adagideli2009,Jacquod2012}. We find helpful for the reader to have, in a nutshell,  the expressions for the charge and spin currents with their explicit dependence on the electron energy and the presence of an external magnetic field. The latter is of particular importance in our study: The magnetic field breaks time-reversal symmetry and drives the symplectic-unitary crossover. The description of the universal spin current fluctuations at the crossover is one of the key results of this paper, as discussed in Sec.~\ref{sec:disorder}.

We consider multi-terminal two-dimensional systems where the electrons flow under the influence of a spin-orbit interaction of the Rashba or/and Dresselhaus type. The device is connected  by ideal point contacts to ${\cal N}$ independent electronic reservoirs, denoted by $i=1,\cdots,{\cal N}$.  The electrodes are subjected to voltages denoted by $V_i$. We use the Landauer-B\"uttiker approach to write the $\alpha$ direction spin resolved current through the $i$th terminal as \cite{Buttiker86}
\begin{equation}
I^{\alpha\sigma}_i = \frac{e^2}{h}\sum_{j,\sigma^\prime}  \sum_{\scriptsize{\small m\in i}\atop{n\in j}} | S_{m,\sigma;n,\sigma^\prime}^{\alpha}|^2 (V_i - V_j)
\end{equation}
where $\sigma$ and $\sigma^\prime$ are the spin projections in the $\alpha=x,y$ or $z$ direction and $S$ is the quaternionic scattering matrix of order $2N_T\times 2N_T$ that describes the transport of electrons through the system. The total number of open orbital scattering channels is  $N_T=\sum^{\cal N}_{i=1}N_i$, where $N_i$ is the number of open channels in $i$th lead point contact.

The electric current at the $i$th terminal is $I^{(0)}_i=I_{i}^{\alpha\uparrow}+I_{i}^{\alpha\downarrow}$, for any direction $\alpha=x,y,z$ of the electron spin projection. Similarly, the $\alpha$-axis component of the spin current $I_i^{(\alpha)}$ is defined as the difference between the two spin projections along the $\alpha$-axis, namely, $I_{i}^{(\alpha)}=I_{i}^{\alpha\uparrow}-I_{i}^{\alpha\downarrow}$.

Let us consider a set up with ${\cal N}=4$ terminals. An applied bias voltage between electrodes $i=1$ and 2 gives rise to a longitudinal electronic current $I$ and due to the spin Hall effect to spin currents at the transversal contacts. \cite{simulacao}   Charge conservation imposes that $I\equiv I_{1}^{(0)}=-I_{2}^{(0)}$. Moreover, due to the absence of a transversal voltage bias, $I_i^{(0)}=0$ for $i=3,4$. Using these constraints, it was shown \cite{jacquod1} that the transversal spin currents are given by
\begin{equation}
J_{i}^{(\alpha)}=\frac{1}{2}\left(\mathcal{T}_{i2}^{(\alpha)}-\mathcal{T}_{i1}^{(\alpha)}\right)-\mathcal{T}_{i3}^{(\alpha)} \widetilde{V}_{3}- \mathcal{T}_{i4}^{(\alpha)} \widetilde{V}_{4}
\label{eq:spincurrent}
\end{equation}
where $i=3,4$. Likewise, the longitudinal charge current reads
\begin{align}
J^{(0)}_i=&\,\frac{1}{4}\! \left(2N_{1}-\mathcal{T}_{11}^{(0)}+2N_{2}-\mathcal{T}_{22}^{(0)}+\mathcal{T}_{12}^{(0)}+\mathcal{T}_{21}^{(0)}\right)\nonumber\\& -\frac{1}{2}\!\left(\mathcal{T}_{23}^{(0)}-\mathcal{T}_{13}^{(0)}\right)\widetilde{V}_{3} -\frac{1}{2}\!\left(\mathcal{T}_{24}^{(0)}-\mathcal{T}_{14}^{(0)}\right)\widetilde{V}_{4} \;,
\label{eq:chargecurrent}
\end{align}
where $i=1,2$. For notational convenience one introduces the dimensionless currents $J=h/e^2(I/V)$. The effective  voltages $\widetilde{V}_{i}$  (in units of $V$), are given by rather lengthy expressions of the generalized transmission coefficients $\mathcal{T}_{ij}^{(0)}$, which can be found in Ref.~\onlinecite{jacquod1}. Finally, the generalized transmission coefficients read
\begin{align}
\mathcal{T}_{ij}^{(\alpha)}(E,E'\!,B,B')={\rm tr}\!\left[\left(1_i \otimes \sigma^\alpha\right)  S^{\dagger}(E,B)1_j S(E'\!,B')\right],
\end{align}
where $E$ and $B$ stand for the electron energy and the magnitude of an external magnetic field.
 The trace is taken over the scattering channels that belong to the $i$th and $j$th point contact. The matrix $1_i$ stands for a projector of the scattering amplitudes into $i$th point contact channels. The matrices $ \sigma^\alpha$,  with $\alpha \in \{x,y,z \} $, are the Pauli matrices, while $ \sigma^0 $ is the $ 2 \times 2 $ identity matrix characterizing an unpolarized (charge) transport.

\section{Universal spin-Hall conductance fluctuations}
\label{sec:disorder}

We assume that the electron dynamics in the quantum dot is ballistic and ergodic, \cite{Beenakker97,Alhassid00} and model the system statistical properties using the random matrix theory. In this section, we describe the procedure to obtain universal  spin and charge current-current autocorrelation functions given by Eqs.~\eqref{eq:spincurrent} and \eqref{eq:chargecurrent}.  We compare our analytical expressions the with results from numerical simulations.

Let us assume that spin-orbit interaction is sufficiently strong to fully break the system spin rotational symmetry. In other words, the  spin-orbit scattering time is much smaller than the electron dwell time in the system, that is, $\tau_{\rm s.o.} \ll \tau_{\rm dwell}$. Hence, in the absence of an external magnetic field, the  scattering matrix has symplectic symmetry. By increasing $B$, the system is driven through a crossover from the symplectic to the unitary symmetry.

The resonance $S$-matrix can be parameterized as \cite{brouwer1}
\begin{eqnarray}
S(E,B)=TU \! \left[1-Q^\dagger R(E,B) Q U\right]^{-1} \!T^\dagger,
\label{SMatriz}
\end{eqnarray}
where $U$ is a matrix of order $2M \times  2M $ of quaternionic form. \cite{Mehta91} $M$ stands for the number of resonances of the quantum dot. We take $M \gg N_T$.  The matrices $ Q $ and $ T $ describe projector operators of order $(2M-2N_T) \times 2M$ and $2N_T \times 2M$, respectively. Their matrix elements read $ Q_{i, j} = \delta_{i +2 N_T, j} $ and $ T_{i, j} = \delta_{i, j} $. The matrix $R$ of order $ (2M-2N_T) \times (2M-2N_T) $ reads
\begin{eqnarray}
R(\epsilon,x)=\exp\left(\frac{i\epsilon}{M} \sigma^0+\frac{x}{M}X\sigma^0 \right) ,
\label{stube}
\end{eqnarray}
where  $X$ is an anti-hermitian Gaussian distributed random matrix, while $\epsilon = (\tau_{\rm dwell}/ \hbar)E$ and $x^2 = \tau_{\rm dwell}/\tau_ {B}$ are dimensionless parameters representing $E$ and $B$.

Both the dwell time $\tau_{\rm dwell}$ and the magnetic scattering time $\tau_B$ are a system specific quantities. $\tau_{\rm dwell}$ is usually expressed in terms of the decay width, namely $\Gamma = \hbar/\tau_{\rm dwell}= N_T\Delta /2\pi$, where $\Delta$ is the system mean level spacing. In turn, $\tau_B^{-1}$ is the rate by which the electron trajectory accumulates magnetic flux in the quantum dot. For chaotic systems,
$\tau_B^{-1}=\kappa ({\cal A} B)/\phi_0$, where $\phi_0$ is the unit flux quantum, ${\cal A}$ is the quantum dot lithographic area, and $\kappa$ is a diffusion coefficient that depends on the quantum dot geometry. \cite{Pluhar95}

The $ S $-matrix given by Eq.~\eqref{SMatriz} can be formally expanded in powers of $U$, namely,
\begin{equation}
S(\epsilon,x)=\sum_{m=0}^{\infty}TU\left[Q^\dagger R(\epsilon,x) Q U\right]^{m}T^\dagger,\nonumber
\end{equation}
and used to calculate the ensemble average of the transmission coefficients $\langle \mathcal{T}_{ij}^{(\alpha)}\rangle$ for  $M \gg N_{T}$
Following this algebraic procedure, we write the sample transmission coefficient as a two-point function of the $S$-matrix,
\begin{align}
\mathcal{T}_{ij}^{(\alpha)}(\epsilon,\epsilon',x,x')=&\!\! \sum_{m,n=0}^{\infty} \!\!{\rm Tr}\!\left\{\left(1_i \otimes \sigma^\alpha\right)T
\left[U^{\dagger}Q^\dagger R^{\dagger}\!(\epsilon,x) Q \right]^{m}
\right.\nonumber\\
&\hskip-0.3cm\times\!\left.
U^\dagger T 1_jTU
\left[Q^\dagger R(\epsilon',x') Q U\right]^{n}T^\dagger\right\}.
\label{Tij}
\end{align}

For chaotic quantum dots, as standard,   \cite{brouwer2} we assume the matrix elements of $U$ as Gaussian variables, with zero mean and variance $1/M$. This allows one to express the calculation of moments and cumulants of $\mathcal{T}_{ij}^{(\alpha)}$ by an integration over the unitary group leading to a diagrammatic expansion in powers of $N_T^{-1}$ in terms of diffusons (ladder) and cooperons (maximally crossed) diagrams. The method is described  for the Dyson ensembles in Ref.~\onlinecite{brouwer1}. This approach has been extended to treat the crossover between symmetry classes \cite{brouwer2,GeneralCrossover} and is shown to render the same results as the quantum circuit theory. \cite{GeneralCrossover}

\begin{figure}[h]
\begin{center}
\includegraphics[width=\columnwidth]{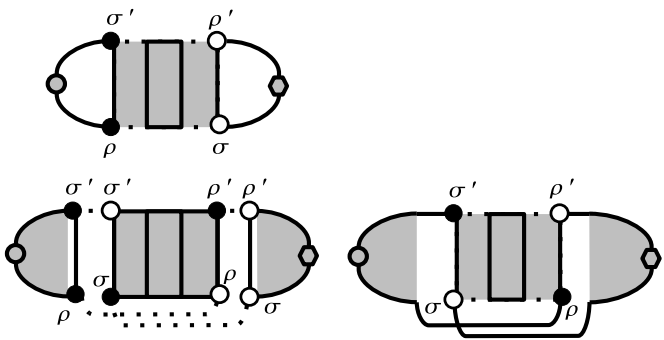}
\end{center}
\vskip-0.5cm
\caption{Diagrams representing the leading contributions to the average transmission coefficient $\langle \mathcal{T}_{ij}^{(\alpha)}\rangle$ of Eq. \eqref{Tij}. Vertical lines represent contractions of the $U$ matrix elements. The white and black dots in  stand for the channel indices of the matrix $U$ and the Greek symbols represent the Pauli matrices indices $\sigma_{\rho,\sigma}$ in the spin space. The first diagram (top) is known as diffuson and accounts for the semiclassical diffusive processes. The two others diagrams are known as cooperons and give the mean quantum correction, namely, the weak (anti)localization corrections.}
\label{fig:diagrams}
\end{figure}

The diagrammatic technique \cite{brouwer1} allows one to calculate moments up to arbitrary order of the transmission coefficients. It consists in grouping and integrating over the Haar measure all the independent powers of $U$ in Eq.~\eqref{Tij}.
Figure \ref{fig:diagrams} shows the diagrams that represent the leading order contributions to all possible contractions of the $U$ matrices.
 In the sum of Eq. \eqref{Tij}, we verify that, after taking the average over $U$, only the powers with $m=n$ contribute to the average transmission coefficients.
The white and black dots in Fig. \ref{fig:diagrams} stand for the indices of the matrix $U$, with elements $U_{ij}$ in the channels space, while the Greek symbols represent the Pauli matrices indices $\sigma_{\rho,\sigma}$ in the spin space.

The diagram at the top row of Fig.~\ref{fig:diagrams} is called diffuson, in analogy to the ladder diagram that appears in diagrammatic expansion to calculate the conductivity in disordered diffusive mesoscopic systems. \cite{mesoscopic} The diffuson contribution to the average transmission coefficient reads
\begin{align}
\langle\mathcal{T}_{ij}^{(\alpha)}(\epsilon,\epsilon',x,x')\rangle^{({\rm d})} =&\frac{1}{2}\sum_{\rho\sigma}\Big[\text{Tr}\left( 1_i\otimes \sigma^\alpha\right)\mathcal{D}\text{Tr}\left(1_j\right)\nonumber\\
&-\text{Tr}\left( 1_i\otimes \sigma^\alpha\right)\mathcal{D}^2\text{Tr}\left(1_j\right)\Big]_{\rho\sigma;\sigma\rho},
\label{Tijm}
\end{align}
where, for clarity, we make explicit the spin degree freedom of the symplectic structure. We use the properties $\text{Tr}\left( 1_i\otimes \sigma^\alpha\right)=2N_i \delta_{\alpha 0}$ and $\text{Tr}\left(1_j\right)=2N_j$  to write
 \begin{eqnarray}
\mathcal{D}^{-1}= 2 M \sigma^0 \otimes\sigma^0 - \text{Tr} \left(R \otimes R'^\dagger\right)
\label{eq:defD}
\end{eqnarray}
with
\begin{align}
\label{eq:defR}
\text{Tr} &\left(R \otimes R'^\dagger\right) =  \sigma^0 \otimes\sigma^0 \\
&\;\;\;\;\times \left[2M- 2 N_T-2i(\epsilon-\epsilon')  - (x-x')^2\right], \nonumber
\end{align}
where $R'$ is a shorthand notation for $R(\epsilon',x')$. The tensor products follow the  ``backward algebra" , \cite{brouwer2,Cremers03} namely, $( \sigma^i \otimes \sigma^j)( \sigma^k \otimes \sigma^l)=  (\sigma^i\sigma^k) \otimes (\sigma^l\sigma^j)$.

The diffuson contribution to the generalized transmission coefficient is obtained by evaluating Eq.~\eqref{Tijm} using the expressions \eqref{eq:defD} and \eqref{eq:defR}. It reads
\begin{align}
\langle\mathcal{T}_{ij}^{(\alpha)}(\epsilon,\epsilon',x,x')\rangle^{({\rm d})} = \delta_{\alpha 0}\frac{N_iN_j}{N_D} \left(2+\frac{1}{N_D}\right),
\end{align}
where $N_D=N_T[1- i\left(\epsilon'-\epsilon\right)+(x'-x)^2/2]$.

We are now ready to evaluate the two maximally crossed diagrams at the bottom row of Fig.~\ref{fig:diagrams}. They are known as cooperons \cite{mesoscopic} and represent the main quantum interference correction to the conductance in chaotic systems, responsible for the weak localization peak.

Following the procedure described above, we obtain the cooperon contribution for the generalized average average transmission, namely,
\begin{align}
\langle\mathcal{T}_{ij}^{(\alpha)}(\epsilon,&\epsilon', x,x')\rangle^{\rm (c)} =\,\frac{1}{2}\sum_{\rho\sigma} \bigg[\,
\text{Tr}\Big(F_i^{(\alpha)}(\mathcal{T}f_{TT}\mathcal{T})F_j\Big)   \nonumber\\
& -\frac{1}{(2M)^3}{\rm Tr}\Big(F_i^{(\alpha)}(\mathcal{T}f_{TT}\mathcal{T})\Big){\rm Tr}(F_j) \bigg]_{\rho\sigma;\rho\sigma}.
\label{Tijmd}
\end{align}
The operator $\mathcal{T} = \sigma^0 \otimes \sigma^y$ is related with the time-reversed of path in the cooperon channel of the dual space. We also define the matrices
\begin{align}
F_i ^{(\alpha)}=&\, 1_i \otimes \sigma^{\alpha}+\text{Tr}(1_i \otimes \sigma^{\alpha}) \mathcal{D} \,(R'^\dagger\otimes R), \nonumber \\
F_j =&\, 1_j +\text{Tr}(1_j) \mathcal{D}\, (R'^\dagger\otimes R),
\end{align}
where
\begin{align}
f_{UU}=&\left[2M\sigma^0\otimes\sigma^0-\text{Tr}(R\otimes R^*)\right]^{-1},\nonumber\\
f_{TT}=&\,(2M\sigma^0\otimes\sigma^0)\,{\rm Tr}(R\otimes R'^*)f_{UU}.
\end{align}
The superscript $*$ denotes the quaternion complex conjugation. Using the quaternionic conjugation rules and taking the limit $M \to \infty$, we obtain
\begin{align}
f_{UU}^{-1}=& \, 2 N_C \sigma^0 \otimes\sigma^0.\nonumber\\
f_{TT}=&\,(2 M  \sigma_0 \otimes \sigma_0)[(2 M - 2N_C) \sigma_0 \otimes \sigma_0] f_{UU},
\end{align}
where $N_C=N_T[1 - i\left(\epsilon'-\epsilon\right)+(x'+x)^2/2]$.

Summing up the diffuson and the cooperon contributions to the generalized transmission coefficients we obtain
\begin{align}
\langle\mathcal{T}_{ij}^{(\alpha)}(\epsilon,\epsilon',x,x')\rangle &=\delta_{\alpha 0}\left\{\frac{2N_iN_j}{N_D} + \frac{N_i}{N_C} \right.\nonumber\\
&\times \left.\left[ \frac{N_j\left[N_D+i (\epsilon-\epsilon')\right]}{N_D^2}-\delta_{ij}\right]\right\}.
\label{condg}
\end{align}
For $\alpha=0$, Eq.\ \eqref{condg} reproduces the average electron transmission reported in the literature. \cite{brouwer1}. As expected in this case, the average spin transmission coefficients are zero.

Let us now analyze the transmission coefficient fluctuations. We use the same diagramatic procedure described above for all the $32$ diagrams characteristic of the usual covariance calculations. \cite{brouwer1} To address relevant physical situations, it is sufficient to consider transmission coefficients with single energy and magnetic field arguments, that is, $\mathcal{T}_{ij}^{(\alpha)}(\epsilon,x)\equiv\mathcal{T}_{ij}^{(\alpha)}(\epsilon,\epsilon,x,x)$. Following the diagrammatic approach, \cite{GeneralCrossover} we calculate the covariance
${\rm cov}[\mathcal{T}_{ij}^{(\alpha)}(\epsilon,x),\mathcal{T}_{kl}^{(\beta)}(\epsilon',x')] \equiv \langle\mathcal{T}_{ij}^{(\alpha)}(\epsilon,x)\mathcal{T}_{kl}^{(\beta)}(\epsilon',x')\rangle- \langle\mathcal{T}_{ij}^{(\alpha)}(\epsilon,x)\rangle \langle\mathcal{T}_{kl}^{(\beta)}(\epsilon',x')\rangle $ and obtain
\begin{widetext}
\begin{eqnarray}
{\rm cov}\!\left[\mathcal{T}_{ij}^{(\alpha)}(\epsilon,x),\mathcal{T}_{kl}^{(\beta)}(\epsilon',x')\right] &=& \delta_{\alpha 0}\frac{N_iN_jN_kN_l}{N_T^2}\left(\frac{1}{\left|N_D\right|^2}+\frac{1}{\left|N_C\right|^2}\right)
+\delta_{\alpha\beta}\frac{\delta_{ik}\delta_{jl}N_iN_j}{\left|N_D\right|^2}+\delta_{\alpha 0}\frac{\delta_{il}\delta_{jk}N_iN_k}{\left|N_C\right|^2}
\nonumber\\
&-&\frac{N_iN_k}{N_T}\left(\delta_{\alpha 0}\frac{\delta_{jl}N_j}{\left|N_D\right|^2}+\delta_{\alpha\beta}\frac{\delta_{ik}N_jN_l}{N_k\left|N_D\right|^2}+\delta_{\alpha 0}\frac{\delta_{jk}N_l+\delta_{il}N_j}{\left|N_C\right|^2}\right)
\label{covg}
\end{eqnarray}
\end{widetext}

\begin{figure}[ht]
\begin{center}
\includegraphics[width=0.95\columnwidth]{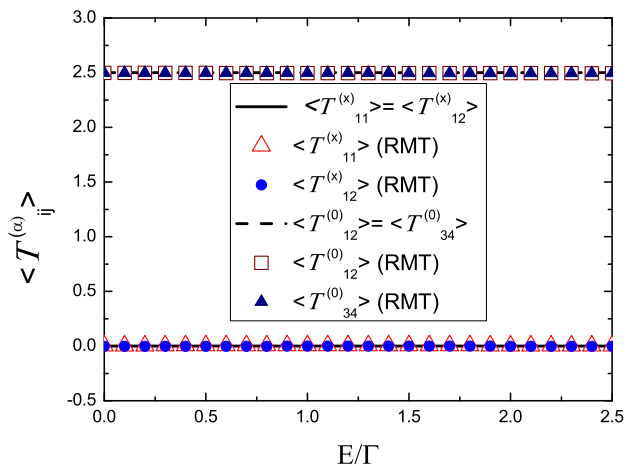}
\end{center}
\vskip-0.5cm
\caption{(Color online) Generalized average transmission coefficient  $\langle \mathcal{T}_{ij}^{(\alpha)}(\epsilon,x=0)\rangle $ versus energy $\epsilon = E/\Gamma$  for different spin and terminal indices. The analytical results (stub model) are represented by the solid and doted lines, while the results of numerical simulations (RMT Hamiltonian model) are represented by the symbols. The figure inset describes the different $\mathcal{T}_{ij}^{(\alpha)}$.   }
\label{fig:THall}
\end{figure}

Support to our analytical findings is provided by numerical simulations. For that purpose, we find convenient to employ the Hamiltonian approach to the $S$-matrix, \cite{mahaux69}. The latter is more amenable for numerical simulations than the $S$ matrix parametrization of Eq.~\eqref{SMatriz} and both are statistically equivalent. \cite{Lewenkopf91}

The Hamiltonian parametrization of the $S$ matrix reads
\begin{equation}
\label{eq:SHeidelberg}
S (E,X) = \mbox{$\openone$} - 2\pi i W^\dagger (E - H(B) + i \pi W W^\dagger)^{-1} W \;,
\end{equation}
where $E$ is the electron propagation energy and $H(B)$ is the matrix of dimension $2 M \times 2 M$ that describes the resonant states ($M$ orbital states times the 2 spin projections). In general, $H$ depends on one (or more) external parameters $X$. As discussed before, we are interested in the case where by increasing $B$ one breaks time-reversal symmetry, driving the system from the symplectic to the unitary symmetry. In our numerically simulations, we consider $B \gg B_c$, namely, the specific case of pure ensembles. Accordingly, $H$ is taken as a member of the Gaussian unitary ensemble corresponding to the case of broken time-reversal symmetric case, usually denoted by $\beta=2$. The matrix $W$ of dimension $M \times (2N_T)$ contains the channel-resonance coupling matrix elements. Since the $H$ matrix is statistically invariant under unitary transformations, the statistical properties of $S$ depend only on the mean resonance spacing $\Delta$, determined by $H$, and $W^\dagger W$. We assume a perfect coupling between channels and resonances, which corresponds to maximizing the average transmission following a procedure described in Ref.\ \onlinecite{vwz85}.

\begin{figure}[h]
\vskip0.3cm
\begin{center}
\includegraphics[width=0.95\columnwidth]{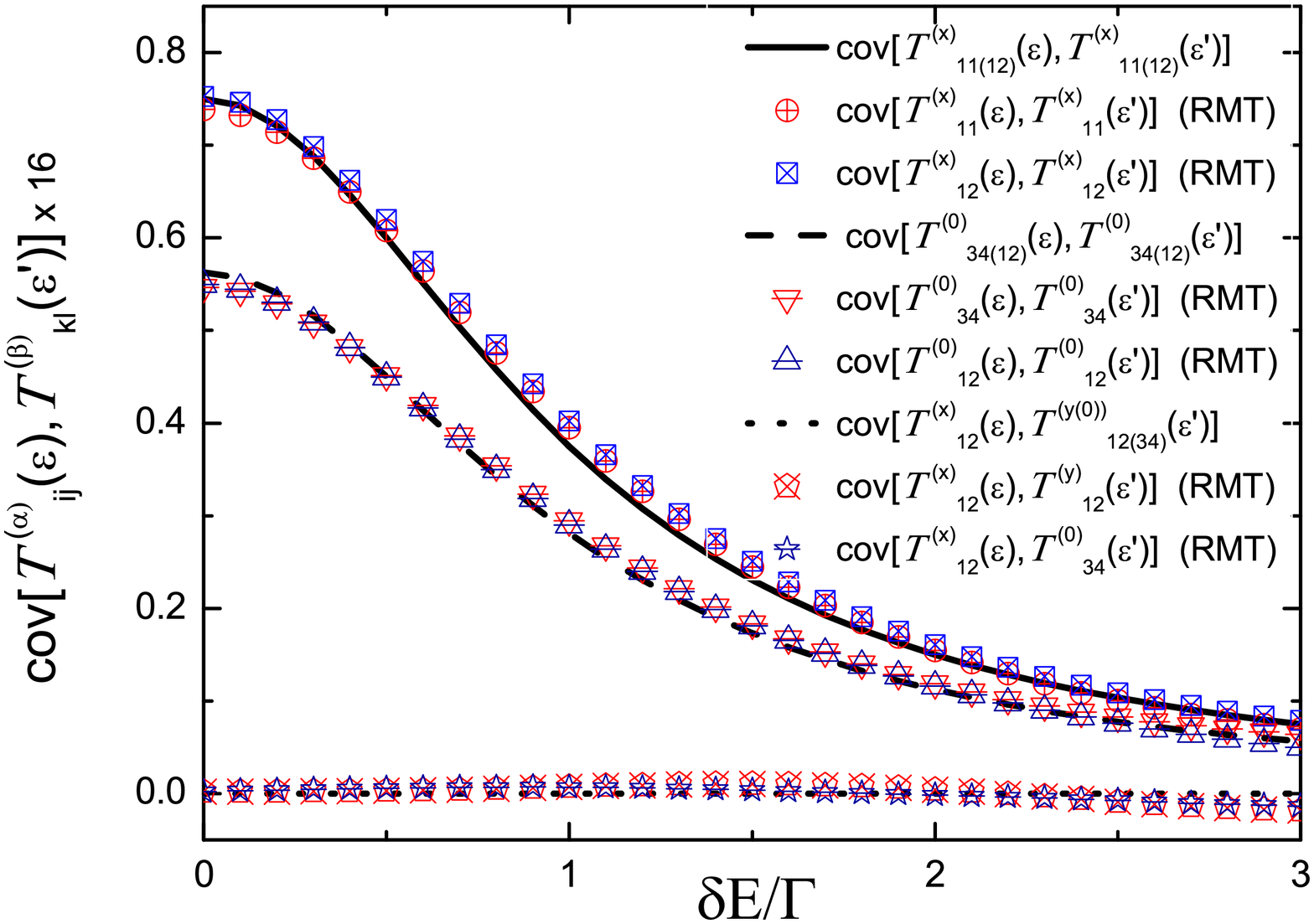}
\end{center}
\vskip-0.5cm
\caption{(Color online) Transmission coefficient covariance ${\rm cov}[\mathcal{T}_{ij}^{(\alpha)}(\epsilon,x=0)\mathcal{T}_{kl}^{(\beta)}(\epsilon^\prime,x=0)] $ as a function of the energy difference $\Delta\epsilon = (E-E^\prime)/\Gamma$  for different spin and terminal indices. The analytical results (stub model) are represented by the solid and doted lines, while the results of numerical simulations (RMT Hamiltonian model) are represented by the symbols. The figure inset describes the different ${\rm cov}[\mathcal{T}_{ij}^{(\alpha)}(\epsilon)\mathcal{T}_{kl}^{(\beta)}(\epsilon^\prime)]$ considered.   }
\label{fig:covT}
\end{figure}

For simplicity, we take the case of $N\equiv N_i$. The results presented in Figs.~\ref{fig:THall} and \ref{fig:covT} correspond to the systems with $N=5$ perfectly coupled modes and $M=400$ resonant levels. Hence, the $S$-matrix has  $2N_T =40$ open channels.  The ensemble averages are taken over  $N_{\rm r} = 10^5$ realizations within an energy interval around the band center, comprising about $M/4$ resonances.

Figure \ref{fig:THall} compares the average transmission $\langle T_{ij}^{(\alpha)}(\epsilon)\rangle$ obtained from the numerical simulations with the analytical expression \eqref{condg} for a number of different cases. The agreement is very good, with accuracy of the order of $N_{\rm r}^{-1/2}$. The simulations indicate that the average transmission in stationary in $\epsilon=E/\Gamma$, as it should. 

Figure \ref{fig:covT} contrasts transmission coefficient covariances calculated using Eq.÷\eqref{covg} with those obtained from numerical simulations for a number of different cases. As before, the discrepancies are very small and stay within the statistical precision $N_{\rm r}^{-1/2}$.  The random matrix theory \cite{vwz85} predicts an autocorrelation length $\Gamma = N_T \Delta/(2\pi)$ for a two-terminal geometry. Our results for the correlation function extend the latter to four-terminal geometries with (or without) spin polarization.

Let us now return to the problem of spin and charge current and effective potential. As mentioned, both are combinations of transmission coefficients. Fortunately, it is possible to calculate average currents and current-current correlation functions in terms of the average and the transmission coefficients correlation functions already calculated and confirmed numerically. The effective voltages $\widetilde{V}_{3}$ and $\widetilde{V}_{4}$ show sample-to-sample fluctuations that depend both on the energy and magnetic field. On the other hand, as discussed in Ref.~\onlinecite{jacquod1}, their ensemble averages depend only on the number of open channels, namely, $ \langle \widetilde{V}_i (\epsilon, x) \rangle = 1 / 2 (N_1-N_2) / (N_1 + N_2) $, with $ i =  3,4$. We also note that the ensemble average of the spin current is always zero, $\langle J^{(\alpha)}_{i}(\epsilon,x) \rangle = 0$, with $i=3,4$ and $\alpha\neq 0$, independently of the energy and magnetic field.

The USCF do not depend on the device geometry (nor on the positions of the terminals), but rather on the number of open channels at each terminal. Without loss of generality, let us analyze the spin current covariance for the case $N_1=N_2=N$ and $N_3=N_4=nN$, a setting that is easily realized in experiments. Here, $n$ is a real positive number that we call ``channel factor". \cite{assimetria} 

For the spin currents, for which $\alpha\neq 0$, we obtain
\begin{align}
{\rm cov}\!\left[ J^{(\alpha)}_i(\epsilon, x),J^{(\alpha)}_i(\epsilon', x')\right] = \frac{1}{8}\frac{n/(1+n)^2}{(1+\delta x^2)^2+\delta \epsilon^2}
\label{covspin}
\end{align}
where  $\delta \epsilon = \epsilon - \epsilon'$ and $\delta x = x - x'$.
It is worth noticing that, for $n=1$ and $\delta \epsilon=\delta x = 0$, Eq.~\eqref{covspin} perfectly reproduces the recent reported results  \cite{simulacao,jacquod1} for the universal fluctuations of the transverse spin conductance, namely, $\textrm{rms}[G_{\rm sH}]=e/4 \pi \{{\rm cov}[ J^{(\alpha)}_i(\epsilon,x),J^{(\alpha)}_i(\epsilon,x)]\}^{1/2} \approx 0.18 e/4 \pi$.

Equation \eqref{covspin} shows that the spin current correlation functions do not depend on the cooperon channels, that give rise to terms containing $N_C$. Hence, these quantities do not depend on the magnetic field, represented by $x$, but rather on its variations, $\delta x$. As a consequence, in the set up we consider, the spin current fluctuations are invariant in the symplectic-unitary crossover regime, a quite remarkable property.

The charge current fluctuations, on the other hand, depend both on the cooperon and diffuson channels, leading to
\begin{widetext}
\begin{align}
{\rm cov}[ J^{(0)}_i(\epsilon,x),J^{(0)}_i(\epsilon',x')] = \frac{1}{16}\!
\left\{\frac{1+2n}{(1+n)^3}\frac{1}{\left(1+\delta x^2\right)^2+\delta \epsilon^2}
+\frac{1}{(1+n)^2}\frac{1}{[1+( 2 x+\delta x)^2]^2+\delta \epsilon^2}\right\},\label{covcarga}
\end{align}
\end{widetext}
where $i=1,2$. The magnetic field, represented by $x$, drives the symplectic-unitary crossover. For $x=0$, one recovers the symplectic limit, while the unitary one is attained when $x\gg 1$. Note that in the absence of ``transverse" leads, or $n=0$, Eq.\ \eqref{covcarga} reproduces the two-terminal result found in Ref.~\onlinecite{brouwer2}.

From Eq.~\eqref{covcarga} it follows that
\begin{equation}
{\rm cov} \{[J^{(0)}_i (\epsilon,x)-J^{(0)}_i(\epsilon,- x)]^{2}\} = \frac{16nx (1 +2 x^2)}{(1 + n) (1 +4 x^2)^ 2},
\end{equation}
which demonstrates that, except for the two-terminal case where $n \neq 0$, the charge currents are not even functions of the magnetic field $x$. \cite{Buttiker86}

\section{Alternative statistical measures}
\label{sec:results}

Equations \eqref{covspin} and \eqref{covcarga} are the main results of this paper. Unfortunately, the statistical sampling required to confirm our predictions for the dimensionless currents is rather large, making the experimental requirements quite daunting.  An easier accessible statistical measure has been recently proposed: \cite{densidade2} The dimensionless current $J_i^{(\alpha)}$ fluctuates as $\epsilon$ and $x$ are varied. Let us call the external parameter $z$. Useful statistical information can be extracted from the number of maxima (or minima) of the $J_i^{(\alpha)}(z)$ in a given interval $[z,z+\delta z]$. Using a scale invariance and maximum entropy principle, we relate the joint probability of  $J_i^{(\alpha)}(z)$ and its derivatives to a general equation for the density of maxima, for spin and/or charge transport. The average densities of maxima, $ \langle \rho_ {z}^{(\alpha)} \rangle$ of the fluctuating current $J^{(\alpha)}_i$ are given by \cite{densidade2}
\begin{align}
\left< \rho_{z}^{(\alpha)} \right> =&\frac{1}{2\pi}\sqrt{\frac{T_4}{T_2}} \label{densidade}\\
T_2=&-\frac{d^2}{d(\delta z)^2}{\rm cov}\!\left[ J^{(\alpha)}_i(\epsilon,x),J^{(\alpha)}_i(\epsilon',x')\right]\bigg|_{\delta z=0}\nonumber\\
T_4=&\frac{d^4}{d(\delta z)^4}{\rm cov}\!\left[ J^{(\alpha)}_i(\epsilon,x),J^{(\alpha)}_i(\epsilon',x')\right]\bigg|_{\delta z=0}\nonumber
\end{align}
where $\delta z$ is $\delta \epsilon$ or $\delta x$.

It is convenient to write the charge current covariance  as a deformed Lorentzian. For parametric variations of $z$, we set ${\rm cov}[ J^{(0)}_i(\epsilon,x),J^{(0)}_i(\epsilon,x')]=\alpha_{z}(n,x)/ [1+(\delta x)^2]^{h_{z}(n,x)}$, where $\alpha_{z}(n,x)$ is a crossover function  and $h_z(n,x)$ characterizes the Lorentzian shape deformation of the charge current correlation as a function of $\delta x$. In terms of the factor $h_{z}$, the average density of maxima reads
\begin{equation}
    \left< \rho_{z}^{(\alpha)} \right>=\frac{1}{2 \pi} \sqrt{6\big[h_{z}(n,x)+1\big]},  \label{densidadeh}
\end{equation}
where $z$ can be either $\epsilon$ of $x$.

\begin{figure}[h!]
\begin{center}
\includegraphics[width=\columnwidth]{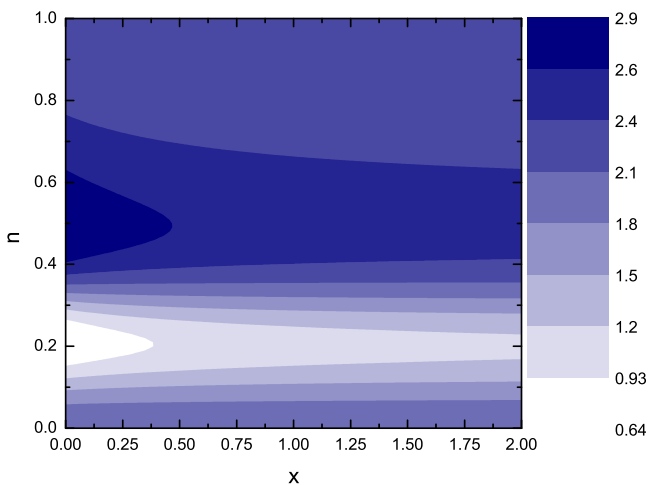}
\end{center}
\vskip-0.5cm
\caption{[Color online] Contour plot of $h_{x}(n, x)$ as a function of the magnetic field, represented by $x$, and the channel factor $n$.  The color code is explained at the strip on the right panel.}
\label{figuramagnetico}
\end{figure}

Using Eqs.~\eqref{covcarga} and \eqref{densidadeh}, we obtain an exact analytical expression for $h_{x}(n,x)$. Its explicit form is not presented here, since it is rather lengthy. Figure~\ref{figuramagnetico} illustrates its general features.  For the unitary symmetry limit, it is well stablished \cite{Efetov,Caio} that the electronic conductance correlation function shows a square Lorentzian behavior Accordingly, we find $h_{\rm sqL}\equiv\lim_{x \rightarrow \infty} h_{x}(n,x)=\lim_{x \rightarrow 0}h_{x}(n,x)=2$, for the pure circular unitary and symplectic ensembles, respectively.  The symplectic-unitary crossover shows a much richer behavior. Figure \ref{figuramagnetico} exhibits a remarkable crossover between sub-Lorentzian, for which $h_{x}<1$, with a minimum value of $h_x \approx 0.64$, and super-Lorentzian, for which $h_{x}>1$ with a maximum value of $h_x \approx 2.92$.

\begin{figure}[h]
\begin{center}
\includegraphics[width=\columnwidth]{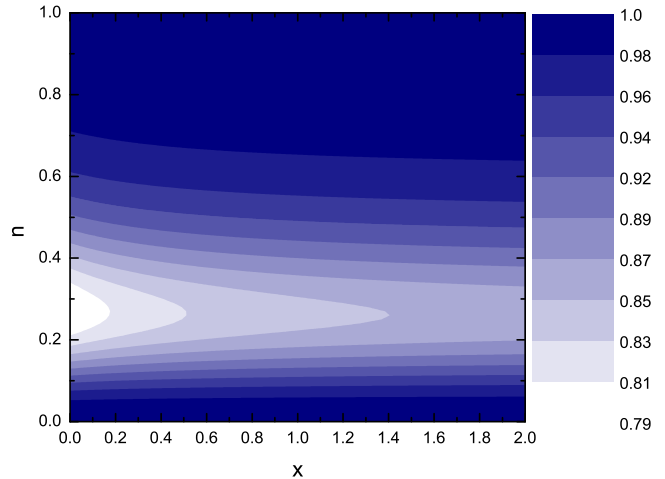}
\end{center}
\vskip-0.5cm
\caption{[Color online] Contour plot of $h_{\epsilon}(n, x)$ as a function of the magnetic field, represented by $x$, and the channel factor $n$.  The color code is explained at the strip on the right panel.}
\label{figuraenergia}
\end{figure}

Parametric variations of $\epsilon$  were first studied in nuclear scattering at low energies and known as Ericson fluctuations \cite{Ericson}. As it is well-known, their characteristic correlation function versus energy has a Lorentzian shape. In the presence of a perpendicular magnetic field and the channel factor, we obtained a unitary-simplectic crossover of Lorentzian-type shapes, generalizing the correlation functions of Ericson fluctuations. For parametric variations of $ \epsilon $, we set ${\rm cov}[ J^{(0)}_i(\epsilon,x),J^{(0)}_i(\epsilon',x)]=\alpha_{\epsilon}(n,x)/ [1+(\delta \epsilon)^2]^{h_{\epsilon}(n,x)}$, where $\alpha_{\epsilon}(n,x)$ is a crossover parameter and $h_{\epsilon}(n,x)$ characterizes the deformation of the Lorentzian shape. As in the previous case, we also obtain a lengthy analytical expression for $h_{\epsilon}(n,x)$. Its main features are displayed in Fig. \ref{figuraenergia}.  As expected, $h_{\rm L} \equiv \lim_{x \rightarrow \infty} h_{\epsilon}(n,x)=\lim_{x \rightarrow 0}h_{\epsilon}(n,x)$=1, for pure circular unitary and symplectic ensembles, respectively. Figure \ref{figuraenergia} exhibits another remarkable crossover from a sub-Lorentzian, for which $h_{\epsilon}<1$, to a Lorentzian behavior.

According to Eq.~\eqref{densidadeh}, the density of maxima corresponding to pure ensembles, namely, $x=0$ or $x \gg 1$, is $ \langle \rho_ {x}^{(\alpha)} \rangle= \sqrt{6 (h_{\rm sqL} +1)} \approx  0.68$ and $\langle \rho_{\epsilon}^{(\alpha)} \rangle= \sqrt{6 (h_{\rm L} +1)} \approx 0.55$ \cite{densidade1,densidade2}, for both spin and charge currents. We emphasize that for the case of the spin correlation function, Eq.~\eqref{covspin}, $ h_{\epsilon} = h_{\rm L} = 1 $ and $h_{x} = h_{\rm sqL} = 2 $ {\it even} in the crossover regime (any value of $ n $ and $x$).

Let us now focus on the longitudinal (charge) correlation function, Eq.~\eqref{covcarga}. For a given value of the channel asymmetry factor, $\langle \rho_x^{(0)}(n,x)\rangle$ has a unique global maximum, $\langle\rho_{x}^{(0)}(n,x_{max}) \rangle$, and minimum, $\langle\rho_{x}^{(0)}(n,x_{min}) \rangle$ The difference, $\Delta \langle \rho_{x} (n)\rangle \equiv \langle\rho_{x}^{(0)}(n,x_{max}) \rangle-\langle\rho_{x}^{(0)}(n,x_{min}) \rangle $ increases with $n$ until it saturates at $ n \approx 3$. In the absence of spin leads, we find the difference $ \Delta \left <\rho_ {x}(0) \right> \approx 0.27$. In the presence of spin leads, we get $ \Delta \left <\rho_{x}(0.5) \right> \approx 0.19$, $\Delta \left<\rho_ {x} (1)\right>  \approx  0.17 $, $\Delta \left< \rho_ {x} (2)\right>  \approx 0.15$, and $\Delta \left< \rho_{x}(5) \right> \approx 0.14$. Thus, in measurements made with a perpendicular magnetic field, for symmetric channels ($n=1$), the spin terminals lead to a reduction in the signal of about $37\%$, which becomes even larger with increasing $n$. Interestingly, the maximum and minimum of $ \left< \rho_{x} \right>$ correspond to magnetic field strengths $ x_{min} \approx 0.20$ and $x_{max} \approx 0.47$, for $ n \in [0,5] $, a rather narrow range of values which is accessible experimentally.

In contrast  to $\left< \rho_{x}(n) \right>$, the energy variation generates a density of conductance peaks containing a single global minimum, $ \left<\rho_ {\epsilon} (n, x_{min})\right> $, and no global maximum. This minimum is located in a very narrow range of values of $ x_{min} \approx 0.26$ as a function of $ n $. The minimum value of the density at these points is of the order of $ \left <\rho_ {\epsilon} \right> \approx 0.52 $.

\section{Summary and Conclusion}
\label{sec:conclusions}

In this paper, we have investigated the spin-Hall conductance fluctuations in a chaotic open quantum dot with spin-orbit interaction. Both the electronic and the spin-Hall conductance fluctuations are universal functions, with autocorrelation functions that depend on the magnitude of the external  magnetic field $B$ and the channel asymmetry factor $n$. A clear intermediate case of symplectic-unitary transitional behavior is found and can be tested experimentally. In particular, the spin current can be measured by using the charge current density of maxima. The results of this Letter  extend the understanding of mesoscopic fluctuations to spin- and charge currents in the symplectic-unitary crossover, characteristic of quantum dots subjected to an external magnetic field.

\acknowledgements
This work is supported in part by the Brazilian funding agencies CAPES, CNPq, FACEPE, FAPERJ, FAPESP, and the Instituto Nacional de Ci\^{e}ncia e Tecnologia de Informa\c{c}\~{a}o Qu\^{a}ntica-MCTI.

\end{document}